\documentstyle{article}

\def\JCP{{\em J. Chem. Phys.}}
\def\CPL{{\em Chem. Phys. Lett.}}
\def\JPC{{\em J. Phys. Chem.}}

\def\JPAM{{\em J. Phys. A, Math.Gen.}}
\def\JMS{{\em J. Mol. Spectr.}}

\def\be{\begin{equation}}
\def\ee{\end{equation}}
\def\bea{\begin{eqnarray}}
\def\eea{\end{eqnarray}}

\begin{document}
\title{Symmetry-Adapted Algebraic Description of Stretching and 
Bending Vibrations of Ozone} 
\author{F. P\'erez-Bernal$^1$, J.M. Arias$^1$,  A. Frank$^{2,3}$, 
 R. Lemus$^2$ and R. Bijker$^2$ 
\and
\begin{tabular}{rl}
$^{1)}$ & Departamento de F\'{\i}sica At\'omica, Molecular y Nuclear,\\
        & Facultad de F\'{\i}sica, Universidad de Sevilla,\\
        & Apdo. 1065, 41080 Sevilla, Espa\~na\\
$^{2)}$ & Instituto de Ciencias Nucleares, U.N.A.M.,\\
        & A.P. 70-543, 04510 M\'exico D.F., M\'exico\\
$^{3)}$ & Instituto de F\'{\i}sica, Laboratorio de Cuernavaca,\\
        & A.P. 139-B, Cuernavaca, Morelos, M\'exico
\end{tabular} }

\date{}
\maketitle

\vspace{6pt}
\begin{abstract}
The vibrational excitations of ozone, including both bending and 
stretching vibrations, are studied in the framework of a symmetry-adapted 
algebraic approach. This method is based on the isomorphism between the 
$U(2)$ algebra and the one-dimensional Morse oscillator, and the 
introduction of point group symmetry techniques. 
The use of symmetry-adapted interactions, which in the harmonic limit have 
a clear physical interpretation, makes it possible to systematically 
include higher order terms and anharmonicities.
A least-square fit to all published experimental levels (up to ten quanta) 
of $^{16}$O$_3$ and $^{18}$O$_3$ yields a r.m.s. deviation of 2.5 and 
1.0 cm$^{-1}$, respectively. 
\end{abstract}
\vspace{10pt}
\begin{center}
Journal of Molecular Spectroscopy, in press\\
Preprint: chem-ph/9605001
\end{center}

\clearpage

\section{Introduction} 

Ozone has been the subject of a large number of studies because of 
its geophysical and atmospheric importance. This molecule has received 
considerable attention from the point of view of spectroscopic 
measurements and theoretical calculations \cite{atlas,exp,chan,barbe}. 
In spite of the high reactivity of the ozone molecule, a rather large 
number of overtone and combination levels has been identified 
\cite{atlas,exp}. These excitations,  involve mostly stretching 
modes, but there is some recent data on  overtones involving the 
fundamental bending mode $\nu_2$. 
The ozone molecule is difficult 
to describe, mainly due to the radically different behaviour displayed 
by the symmetric $\nu_1$ and antisymmetric $\nu_3$ stretching modes. 
Whereas $\nu_1$ shows a nearly harmonic behaviour, $\nu_3$ has large 
anharmonicities.

The spectroscopic constants of ozone have been studied by many authors 
using  {\it ab initio} methods. These methods involve a significant 
computational effort and have been used to calculate the harmonic frequencies 
$\omega_i$ (see e.g. \cite{lee,Watts,Borowski,Peterson})
and the anharmonic constants $X_{ij}$ (see e.g. \cite{lee,Peterson}). 
Whereas the {\it ab initio} calculations of \cite{lee} give reasonable 
results for the $A_1$ modes, the results for the $B_2$ mode (especially 
the anharmonicity constants) still show large 
deviations from experimental values. 
The more recent calculation of 
\cite{Peterson} shows much improved results for the anharmonicity 
constants and includes the Darling-Dennison resonance constant, 
but finds systematically low values for the harmonic frequencies. 
It was also shown that the results for the harmonic frequencies 
can be improved 
by including a Coriolis correction. 
The standard deviation between these 
two calculations (without and with Coriolis correction) and 17 observed 
vibrational energies (up to three quanta and one state corresponding to 
four quanta) is 94.6 and 8.3 cm$^{-1}$, respectively 
(see Table 5 of \cite{Peterson}).

The vibrational energies of ozone have also been analyzed with
algebraic methods.  We  mention the coupled $SU(2)$ methods of 
\cite{chan,roo,ben} 
and the $SU(n)$ approach for a set of $n$ coupled anharmonic
oscillators of \cite{kell1}. These methods are phenomenological and hence
require a minimal set of experimental data to determine the
interaction strenghts.  Since the energies 
and corresponding wave functions are obtained by 
matrix diagonalization, the required computing time is small 
compared to that of integro-differential methods that solve the 
Schr\"odinger equation in configuration space. 
Especially for molecules 
for which the configuration space that has to be included to provide 
reliable {\it ab initio} results becomes very large, algebraic 
methods provide a numerically efficient,  alternative  tool  
to study molecular vibrations. All  four studies 
\cite{chan,roo,ben,kell1} of ozone
were limited to stretching vibrations only.  The results 
 presented in \cite{roo,ben,kell1} 
were obtained by fitting to the same set of 24 vibrational energies and 
show a standard deviation from experiment of 23.7, 9.1 and 7.9 cm$^{-1}$, 
respectively. The results of \cite{chan} which were obtained by applying 
the model of \cite{ben} to a new set of 21 vibrational energies yield a
standard deviation of  3.0 cm$^{-1}$.

In this article we present a study of the experimental vibrational 
spectrum of ozone, including both stretching and bending vibrations, 
in a symmetry-adapted algebraic approach. This  approach 
is based on the use of symmetry-adapted algebraic operators, which in the 
harmonic limit have a direct connection with the configuration space
operators.  
This allows to construct systematically all possible physically meaningful 
interactions up to a certain order \cite{letter,x3}. A least-square fit 
to {\em all} published $^{16}$O$_3$  experimental levels   
(up to ten quanta with a total of 44 stretching and bending vibrations)  
yields a standard deviation of 2.4 cm$^{-1}$. After scaling the boson 
numbers, the harmonic frequencies and the anharmonicities with the mass, 
we find in a (two parameter) fit to 15 vibrational energies of $^{18}$O$_3$ 
a standard deviation of 2.2 cm$^{-1}$. 
In another calculation, in which 
also the harmonic frequencies and the anharmonicities were fitted, we 
find that the standard deviation decreases to 0.8 cm$^{-1}$.
With these parameter sets we calculate the vibrational 
(unobserved) energies of ozone up to ten quanta for $^{16}$O$_3$ 
and up to four quanta for $^{18}$O$_3$. 

The paper is organized as follows. In section 2 we review briefly the main 
aspects of the $U(2)$ algebraic model for a diatomic molecule. In section 3
we discuss the extension of this approach to the ozone molecule. 
In section 4 we introduce symmetry-adapted operators and the harmonic limit,
which are applied in section 5 to the vibrational energies of the ozone 
molecule. Finally, in section 6 we present our summary and conclusions.

\section{The $U(2)$ Algebraic Approach} 

In the framework of the $U(2)$ algebraic approach to diatomic molecules 
a $U(2)$ Lie algebra is associated with the relative coordinate 
connecting the two atoms \cite{vibron,lap}. 
This assignment is based on the isomorphism of the $U(2)$  
algebra and the one-dimensional Morse oscillator \cite{morsis}.
All observables of interest are expressed
in terms of the generators $\hat G = \{ \hat J_{x}, \hat J_{y},
\hat J_{z}, \hat N \}$, which satisfy the commutation relations
\bea
\left[\hat{J}_z,\hat{J}_{\pm}\right]&=&\pm \hat{J}_{\pm} ~,
\nonumber\\
\left[\hat{J}_+,\hat{J}_-\right]&=&2 \hat{J}_z ~,
\nonumber\\
\left[\hat{J}_i,\hat{N}\right]&=& 0 ~,
\label{eq:conmoan}
\eea
where $i=\pm,z$. Here $\hat J_i$ satisfy the `angular momentum' 
commutation relations of $SU(2)$ and $\hat N$ is the number operator.
In particular, the realization of the Morse Hamiltonian in terms of 
$\hat G$ is given by 
\bea
\hat H &=& \frac{A}{4} \left( \hat{N}^2 - 4\hat J^2_z \right)
\;=\; \frac{A}{2} \left( \hat J_+\hat J_- + \hat J_-\hat J_+ - \hat N 
\right) ~.
\label{eq:hmorse2}
\eea
In (\ref{eq:hmorse2}) we have used the relations
\bea
\hat J_z^2 &=& \hat J^2 
- \frac{1}{2}(\hat J_+\hat J_- + \hat J_-\hat J_+) ~,
\nonumber\\
\hat J^2 &=& \frac{1}{4} \hat N (\hat N +2) ~.
\label{jang}
\eea
The eigenstates of $\hat H$ can be labeled by $|j,m\rangle$. 
The last relation of (\ref{jang}) makes it possible to identify $j=N/2$.
The quantum number $m$ takes the values $m=j,j-1,\ldots,-j$, 
which means that the Morse spectrum is reproduced twice. 
Therefore, the values of $m$ are restricted to be nonnegative 
$m=N/2, (N-2)/2, \ldots, 1/2$ or 0 for $N$ odd or even, respectively.
The eigenfunctions are usually written as 
\bea
|[N],v \rangle &=& 
\sqrt{\frac{(N-v)!}{N! v!}} \, (\hat J_{-})^v \, |j,m=j \rangle.
\label{eq:morseigenf2}
\eea
Here $N$ is the total number of bosons fixed by the 
potential shape \cite{vibron,lap} and $v$ corresponds to the number of
quanta in the oscillator. $N$ and $v$ are related to 
the usual labels $j$ and $m$ by 
\bea
N &=& 2 j ~, 
\nonumber\\
v &=& j - m ~.
\label{eq:Nv}
\eea
The energy eigenvalues of (\ref{eq:hmorse2}) take the form
\bea
E &=& \frac{A}{4} ( N^2-4m^2 )
\nonumber\\
&=& -\frac{A(2N+1)}{4} + A(N+1) (v+\frac{1}{2}) 
- A (v+\frac{1}{2})^2 ~ , 
\label{eq:Emorse}
\eea
which correspond  to a Morse-like spectrum. The correspondence can also be 
 established directly by starting from the radial equation of a
two-dimensional 
harmonic oscillator and carrying out a point transformation which
leads to the  
one-dimensional Morse Hamiltonian \cite{morsis}.  This analysis shows
that in the algebraic model harmonic oscillators are transformed into
anharmonic ones by means of the $SU(2)$ algebra \cite{letter}. 

We now turn our attention to the harmonic limit of the model. 
In the harmonic oscillator the number of bound levels is infinite. 
Since according to (\ref{eq:Nv}) $v=0,1,2, \dots, [N/2]$, an infinite 
number of bound states corresponds to taking 
the limit $N\rightarrow \infty$. In order to study the harmonic limit
we scale the parameter $A=A^{\prime}/N$ such that
$A^{\prime}=AN$ remains finite for $N \rightarrow \infty$.
Hence in the harmonic limit the eigenvalues of (\ref{eq:Emorse}) 
reduce to 
\bea
E &\rightarrow& -\frac{A^{\prime}}{2} + A^{\prime} (v + \frac{1}{2}) ~.
\label{eq:Emorse2}
\eea
This shows that in the algebraic approach the harmonic limit is 
obtained by taking $N\rightarrow \infty$. 
 
\section{Algebraic Approach to the Ozone Molecule}

We now present the application of the algebraic model to the case
of the ozone molecule.  As explained in  the previous section a $U^i(2)$
algebra is assigned to each internal coordinate. As shown in Figure 1 
for the ozone molecule, these are associated with the three relative 
coordinates ($i=1,2,3$). The point group symmetry associated with 
the ozone molecule is ${\cal C}_{2v}$.
In this case, however, it is possible to limit ourselves 
to the group ${\cal C}_{2}$ since for a triatomic molecule the three nuclei 
define a plane. On the other hand, the Hamiltonian is expressed in terms 
of operators which are labeled by the interatomic interactions. 
This makes it possible to use the permutation group $S_2$ instead of 
${\cal C}_{2}$. We shall use the isomorphism 
\bea
E &\leftrightarrow& e ~,  
\nonumber\\ 
C_2 &\leftrightarrow& (12) ~.   
\label{eq:c2s2}
\eea
In this article we use the $S_2$ labeling: the $S_2$ group has two 
one-dimensional irreducible representations, which are associated 
with symmetric ($A$) and antisymmetric ($B$) modes, respectively.

A normal mode analysis shows that the ozone molecule exhibits 
three normal modes, 
two of them corresponding to the fundamental stretching modes $(A_1, B_2)$, 
and the other to the fundamental bending mode $(A_1)$ \cite{herzberg}. 
The algebras $U^i(2)$ with $i=1,2$ describe equivalent O-O interactions, 
while the third one with $i=3$ is associated with the bending mode. 
In the latter case  it may be more appropriate to interpret the
$U(2)$ algebra as associated to a P\"oschl-Teller  
potential \cite{vibron,posch}. The molecular dynamical algebra is then 
given by the product
\be
U^{1} (2) \otimes U^{2} (2) \otimes U^{3} (2) ~ , 
\label{eq:da}
\ee
which means that every operator in the model is expressed in terms of 
the generators of the $U^i(2)$ groups \cite{lap}. 
The most general one- and two-body Hamiltonian for the O$_3$ molecule
 which conserves the  number of vibrational quanta, can be written in
terms of three kinds  
of operators: 
$\hat J_{z,i}^2$, representing the noninteracting one-dimensional
anharmonic 
oscillators, plus two types of bond-bond interactions, 
$\hat J_{z,i} \hat J_{z,j}$  
and $\hat J_i \cdot \hat J_j$. For our purposes, it is convenient to 
introduce the combinations 
\bea
\hat H_i &=& \frac{1}{4} \left( \hat N_i^2 - 4\hat J_{z,i}^2 \right)
\;=\; \frac{1}{2} \left( \hat J_{+,i} \hat J_{-,i} 
+ \hat J_{-,i} \hat J_{+,i} - \hat N_i \right) ~,
\nonumber\\
\hat H_{ij} &=& \frac{1}{2} \left( 4\hat J_{z,i} \hat J_{z,j} 
- \hat N_i \hat N_j \right) ~, 
\nonumber\\
\hat V_{ij} & = & 2 \left( \hat J_i \cdot \hat J_j 
- \hat J_{z_i} \hat J_{z_j} \right) 
\;=\; \hat J_{+i}\hat J_{-j} + \hat J_{+j}\hat J_{-i} ~. 
\label{eq:opant}
\eea
The operator $\hat H_i$ corresponds exactly to the Morse Hamiltonian of 
(\ref{eq:hmorse2}). If in addition we impose the invariance under the 
permutation group $S_2$, the Hamiltonian can be expressed in terms of 
the interactions
\bea
\hat H &=& A_1 \, (\hat H_1 + \hat H_2) + A_3 \, \hat H_3 
+ B_{13} \, (\hat H_{13} + \hat H_{23}) + B_{12} \, \hat H_{12} 
\nonumber\\
&& + \lambda_{13} \, (\hat V_{13} + \hat V_{23}) 
+ \lambda_{12} \, \hat V_{12} ~.
\label{eq:halg2}
\eea
Since the interaction terms of (\ref{eq:halg2}) are invariant under the  
$S_2$ group, the corresponding wave functions automatically span the 
irreducible representations of this group \cite{ham}. 

The Hamiltonian of (\ref{eq:halg2}) can be diagonalized in the local 
basis, which is associated to the subgroup chain
\be
\begin{array}{lllllll}
U^{(1)} (2)\otimes&\hspace{-.27cm}U^{(2)} (2)\otimes&\hspace{-.27cm}U^{(3)} 
(2)\supset&\hspace{-.27cm}SO^{(1)} (2) \otimes&\hspace{-.27cm}SO^{(2)} (2)
\otimes&\hspace{-.27cm}SO^{(3)} (2) \supset&\hspace{-.27cm}SO(2) \\ 
~\downarrow&~\downarrow&~\downarrow& \downarrow& \downarrow& \downarrow& 
\downarrow \\
| \; [N_1] &[N_1] &[N_3];&v_1 &v_2 &v_3;&V \; \rangle~, 
\label{eq:lmc}
\end{array}
\ee
\noindent
where below each group we have indicated the quantum numbers characterizing 
the eigenvalue of the corresponding invariant operator. 
The $S_2$ symmetry is reflected in the fact that the total number of 
bosons for the two stretching modes is the same $N_1=N_2$. The number of
bosons for the bending mode is $N_3$, while 
the labels  $v_i$ denote the number of  quanta  
in each oscillator. The total number of quanta  $V=v_1+v_2+v_3$ 
is conserved by the Hamiltonian of (\ref{eq:halg2}). 
Fermi resonances which involve nondiagonal matrix elements in the total 
number of quanta  are not taken into account in the present analysis.
We note that because of the ratio of stretching and bending
frequencies in ozone it may be more appropriate to take $V^\prime =
2v_1 + v_2 + 2v_3$, and to study interactions that are
diagonal in the polyads characterized by these  values of
$V^\prime$ \cite{Hilico}. This can be readily done within the model.  
In the present paper, however, we take a Hamiltonian
that is diagonal in $V$ to analyze the vibrational spectrum of ozone.

The matrix elements of $\hat H_i$, $\hat H_{ij}$ and $\hat V_{ij}$ 
in the local basis of (\ref{eq:lmc}) are given by (ref.)
\bea 
\langle [N_1][N_3]; v_1 v_2 v_3; V| \, \hat H_i \, |
[N_1],[N_3]; v_1 v_2 v_3; V \rangle 
&=& -v^2_i + N_i v_i ~, 
\nonumber\\
\langle [N_1][N_3]; v_1 v_2 v_3; V | \, \hat H_{ij} \, | 
[N_1][N_3]; v_1 v_2 v_3; V \rangle
&=& 2 v_i v_j - (v_i  N_j + v_j N_i ) ~, 
\nonumber\\
\langle [N_1] [N_3]; v^\prime_1 v^\prime_2 v^\prime_3; V | 
\, \hat V_{ij} \, |  [N_1] [N_3]; v_1 v_2 v_3; V \rangle &=&
\nonumber
\eea
\bea
&\hspace{3cm}&\sqrt{ v_j (v_i + 1) (N_i - v_i)(N_j - v_j +1)} 
~\delta_{v^\prime_i,v_i+1} ~ \delta_{v^\prime_j,v_j-1}
\nonumber\\
&&+\sqrt{ v_i (v_j +1) (N_j - v_j) (N_i - v_i +1)} 
~\delta_{v^\prime_i,v_i-1} ~ \delta_{v^\prime_j,v_j+1} ~.  
\label{eq:elma}
\eea
The local mode basis (\ref{eq:lmc}), however,  does not carry the 
irreducible representations of the appropriate point group. 
From this point of view it is more 
convenient to use a symmetry-adapted basis from the beginning. 
This can be done either by projecting the states with $V$ quanta  to the 
various symmetry species directly, 
or by first considering the one-quantum states and then constructing 
the states with higher-quanta  by means of coupling coefficients. 
The latter procedure has the advantage of directly providing 
the symmetry labels \cite{metano}. 
We present here a brief discussion of this procedure 
for the particular case of ozone.
From the one-quantum  local functions
\bea
| 1 \rangle &\equiv& | [N_1], [N_3]; 100; 1 \rangle  ~ , \nonumber\\
| 2 \rangle &\equiv& | [N_1], [N_3]; 010; 1 \rangle  ~ , \nonumber\\
| 3 \rangle &\equiv& | [N_1], [N_3]; 001; 1 \rangle  ~ ,
\label{eq:ops}
\eea
we readily obtain the normalized symmetry-adapted functions 
\bea
 ^1 \psi^{A_s} & = & {1 \over \sqrt{2}} 
\left\{ | 1 \rangle + | 2 \rangle \right\} ~ ,  \nonumber\\ 
 ^1 \psi^{B_s} & = & {1 \over \sqrt{2}} 
\left\{ | 1 \rangle - | 2 \rangle \right\} ~ ,  \nonumber\\ 
 ^1 \psi^{A_b} & = & | 3 \rangle ~,
\label{eq:sab}
\eea
where we have introduced the notation $^V \Psi^\Gamma$ for the wave 
functions. The label $\Gamma$ denotes the point group symmetry.
The states $^1 \psi ^{A_s} $ and $^1 \psi ^{B_s} $ correspond to 
the fundamental symmetric and antisymmetric stretching modes, 
respectively, while the function $^1 \psi ^{A_b}$ corresponds to 
the symmetric bending mode. The higher-quanta  wave functions can 
be obtained through successive couplings \cite{ham}
\bea
^{V_1 + V_2}\Psi^\Gamma_\gamma &=&
\sum_{\gamma_1, \gamma_2} C(\Gamma_1\Gamma_2 \Gamma; \gamma_1 
\gamma_2 \gamma) ~^{V_1}\Psi^{\Gamma_1}_{\gamma_1} ~ 
^{V_2}\Psi^{\Gamma_2}_{\gamma_2} ~.
\eea
In this case the procedure is straightforward, since the 
  Clebsch-Gordan coefficients $C( ; )$ for the $S_2$ group are either 
zero or one. The coupled wave function in (16) corresponds to a total 
number of quanta $V=V_1+V_2$. This approach to construct the function 
space directly provides spectroscopic labels and simplifies the 
diagonalization procedure by splitting the Hamiltonian matrix into blocks, 
which are characterized by their transformation property under the point 
group. For the ozone molecule  the analysis is very simple due to the
absence of  
degenerate modes and the Hamiltonian matrix only splits into two blocks, 
corresponding to the $A$ and $B$ irreducible representations of $S_2$.

We note that in the present application the advantage of using  a
symmetry-adapted basis is limited. However, for molecules with a
higher degree of symmetry, involving degenerate vibrational levels and/or
spurious modes such as for ${\cal D}_{3h}$ triatomic or larger
molecules \cite{letter,x3,metano,ch4}, 
the use of a symmetry-adapted basis becomes essential. 

\section{Symmetry-Adapted Operators and Harmonic Limit}

The interaction terms of (\ref{eq:halg2}) do not  have a specific 
connection with configuration space interactions   and are usually
not convenient  
from a numerical point of view (i.e. in finding the minimum  
in a least-square fit to the data). Both problems can be addressed by 
rewriting the algebraic model interactions in a symmetry-adapted 
form, which in the harmonic limit reduces to definite coordinate-space 
interactions. This connection between algebraic and coordinate-space 
operators provides a physical interpretation of algebraic 
Hamiltonians. The general procedure is discussed in \cite{letter} 
in an application to Be$_4$. 
Applying the same procedure to the interaction terms of (\ref{eq:halg2}) 
we find the linear combinations 
\bea
\hat{\cal H}_{A_s} &=&  
-\frac{1}{2N_1} ( \hat H_{12} - \hat V_{12}) ~,
\nonumber\\
\hat{\cal H}_{B_s} &=&
-\frac{1}{2N_1} ( \hat H_{12} + \hat V_{12}) ~,
\nonumber\\
\hat{\cal H}_{A_b} &=& \frac{1}{N_3} \hat H_3 ~,
\nonumber\\
\hat{\cal H}_{A_sA_b} &=& 
\frac{1}{\sqrt{2N_1N_3}} (\hat V_{13} + \hat V_{23} ) ~,
\nonumber\\
\hat{\cal V}_{1}  &=& 
\frac{1}{N_1} ( \hat H_1 + \hat H_2 + \hat H_{12} ) ~,
\nonumber\\
\hat{\cal V}_{2} &=& \frac{1}{\sqrt{N_1N_3}}
\left[ \hat H_{13} + \hat H_{23} 
+ \frac{N_3}{N_1} ( \hat H_1 + \hat H_2 ) 
+ 2\frac{N_1}{N_3} \hat H_3 \right] ~.
\label{eq:parh}
\eea
Just as for the Hamiltonian of (\ref{eq:halg2}), these interaction terms
are all scalars under $S_2$. We now rewrite the algebraic Hamiltonian 
of (\ref{eq:halg2}) as 
\bea
\hat{\cal H}&=&\alpha_1 \, \hat{\cal H}_{A_s} +
               \alpha_2 \, \hat{\cal H}_{A_b} +
               \alpha_3 \, \hat{\cal H}_{B_s} +
               \alpha_4 \, \hat{\cal H}_{A_sA_b} +
               \beta_1 \, \hat{\cal V}_{1} +
               \beta_2 \, \hat{\cal V}_{2} ~.
\label{eq:halg3}            
\eea 
The advantage of this symmetry-adapted formulation  becomes more
transparent by considering the harmonic limit $(N \to \infty)$. 

In order to analyze the harmonic limit of the operators appearing 
in (\ref{eq:parh}), we first consider it for the generators of 
$SU(2)$  which satisfy the commutation relations (\ref{eq:conmoan}).
The action of $\hat{J}_{\pm}$ is given by
\bea
\hat{J}_+\mid [N],v \rangle &=& \sqrt{v(N-v+1)}   \mid [N], v-1 \rangle~,
\nonumber\\
\hat{J}_-\mid [N],v \rangle &=& \sqrt{(v+1)(N-v)} \mid [N], v+1 \rangle~,
\label{eq:elgen}
\eea
where $\mid [N],v \rangle$ was defined in (\ref{eq:morseigenf2}).
If we rescale the operators $\hat J_{\pm}$ by $\sqrt{N}$ and take 
the $N \rightarrow \infty$ limit, we find 
\bea
\lim_{N \rightarrow \infty} \frac{\hat J_+}{\sqrt{N}} \mid [N],v \rangle 
&=& \sqrt{v} \mid [N],v-1 \rangle ~,
\nonumber\\
\lim_{N \rightarrow \infty} \frac{\hat J_-}{\sqrt{N}} \mid [N],v \rangle 
&=& \sqrt{v+1} \mid [N],v+1 \rangle ~.
\label{eq:limit}
\eea
The r.h.s. corresponds to the harmonic oscillator relations
\bea
b \, \mid [N],v \rangle &=& \sqrt{v} \mid [N],v-1 \rangle ~,
\nonumber\\
b^{\dagger} \, \mid [N],v \rangle &=& \sqrt{v+1} \mid [N],v+1 \rangle ~.
\label{eq:ho}
\eea
This is in agreement with the role played by $N$ in (\ref{eq:Emorse}) and 
(\ref{eq:Emorse2}), i.e. for infinite potential depth the Morse or
P\"oschl-Teller  potentials  
cannot be distinguished from  harmonic oscillator potentials.  
The commutation relation $[ \hat J_+,\hat J_- ]=2\hat J_z$ can be 
rewritten as 
\bea
\frac{1}{N} \left[ \hat J_+,\hat J_- \right] &=& 1-\frac{2\hat v}{N} ~,
\label{eq:conlimhar}
\eea
where we have used $\hat v = (N-2\hat J_z)/2$ (see Eq.~(\ref{eq:Nv})). 
Hence the harmonic limit ($N\rightarrow\infty$) leads to the contraction 
of the $SU(2)$ algebra to the Weyl algebra generated 
by $b$, $b^\dagger$ and $1$ \cite{Gilmore}, with the 
usual boson commutation relation $\left[b,b^\dagger\right]=1$. 

The procedure to arrive at the harmonic limit of the model is now clear: 
each $\hat{J}_{+,i}$ and $\hat{J}_{-,i}$ is rescaled by 
$\sqrt{N_i}$ before taking the $N_i\rightarrow \infty$ limit. 
We then get
\bea
\lim_{N_i\rightarrow \infty} \frac{\hat{J}_{+,i}}{\sqrt{N_i}}
&=& b_i ~, \nonumber\\
\lim_{N_i\rightarrow \infty} \frac{\hat{J}_{-,i}}{\sqrt{N_i}}
&=& b^\dagger_i ~,
\label{eq:conmoan33}
\eea
where the boson operators $b_i$ and $b^\dagger_j$ satisfy the standard 
commutation relation $[b_i,b^\dagger_j] = \delta_{ij}$~ and are
related to the local coordinates and momenta by the usual relations. 
Applying this procedure to the operators in (\ref{eq:halg3}), we obtain
\bea
\lim_{N_1\rightarrow\infty}\hat{\cal H}_{A_s}&=&
b^\dagger_{A_s} b_{A_s} ~,
\nonumber\\
\lim_{N_1\rightarrow\infty}\hat{\cal H}_{B_s}&=&
b^\dagger_{B_s} b_{B_s} ~,
\nonumber\\
\lim_{N_3\rightarrow\infty}\hat{\cal H}_{A_b}&=&
b^\dagger_{A_b} b_{A_b} ~,
\nonumber\\
\lim_{N_1,N_3\rightarrow\infty}{\cal H}_{A_sA_b}&=& 
b^\dagger_{A_s} b_{A_b} + b^\dagger_{A_b} b_{A_s} ~,
\nonumber\\
\lim_{N_1\rightarrow\infty}\hat{\cal V}_{1}&=& 0 ~,
\nonumber\\
\lim_{N_1,N_3\rightarrow\infty}\hat{\cal V}_{2}&=& 0 ~.
\label{eq:hlim}
\eea
The creation operators $b^\dagger_{\Gamma}$ are in turn  given by
\bea
b^\dagger_{A_s}&=&\frac{1}{\sqrt{2}}(b_1^\dagger+b_2^\dagger) ~,
\nonumber\\
b^\dagger_{B_s}&=&\frac{1}{\sqrt{2}}(b_1^\dagger-b_2^\dagger) ~,
\nonumber\\
b^\dagger_{A_b}&=&b_3^\dagger ~,
\label{eq:nanb2}
\eea
with  similar expresions for the annihilation operators. These 
operators also  satisfy the usual boson commutation relations
$[b_{\Gamma},b^\dagger_{\Gamma^{\prime}}]=\delta_{\Gamma,\Gamma^{\prime}}$~.
In Figs.\ 2 and 3 we show the dependence of the eigenvalues of 
$\hat{\cal H}_{A_s}$ and $\hat{\cal H}_{A_b}$ on $N_1$ and $N_3$, 
while in Fig.\ 4  the results for $\hat{\cal V}_{1}$ for two 
and three quanta are shown. $\hat{\cal H}_{B_s}$ and
$\hat{\cal H}_{A_s}$ 
have the same dependence on $N_1$. In all cases the $N_i$ dependence vanishes 
in the harmonic limit. The behaviour of $\hat{\cal H}_{A_sA_b}$ and 
$\hat{\cal V}_{2}$ is similar, but they have a more complicated 
dependence on both $N_1$ and $N_3$. Relations (\ref{eq:hlim}) allow a 
 comparison between the algebraic model parameters and those obtained 
from force-field calculations, since they can be
readily expressed in terms of the conventional coordinates and
momenta \cite{letter}. 

Besides providing a  connection with coordinate space interactions
in the harmonic limit, the interaction terms 
in (\ref{eq:halg3}) have a definite numerical advantage. In a fitting 
procedure to experimental vibrational energies the convergence is greatly 
improved when we replace the Hamiltonian of (\ref{eq:halg2}) by that 
of (\ref{eq:halg3}), 
although these two Hamiltonians are identical.
In the symmetry-adapted form of (\ref{eq:halg3}), each one of the 
interaction terms has a selective action on the fundamental modes, 
whereas the interaction terms of (\ref{eq:halg2}) are linear 
combinations of those of (\ref{eq:halg3}), and hence effect 
all modes simultaneously. This is the reason why the use of the 
symmetry-adapted form of the Hamiltonian improves the numerical 
convergence of the fitting procedure. 

It is also possible to consider 
the relations (\ref{eq:conmoan33}) in the opposite sense. Given a function 
of the harmonic operators $b_i$ and $b_i^\dagger$, its anharmonic 
representation can be found by substitution by the operators 
$\hat J_+/\sqrt{N}$ and $\hat J_-/\sqrt{N}$, respectively \cite{letter}.
This provides a systematic procedure to construct an algebraic 
(anharmonic) realization of a given operator in configuration space.
Again, for the ozone molecule the symmetrization of operators in
rather trivial, but for larger molecules, such as Be$_4$ \cite{letter} 
and CH$_4$ \cite{ch4}, this approach leads to new interactions with a 
specific physical interpretation. 
 
\section{Results for the $^{16}$O$_3$ and $^{18}$O$_3$ Ozone Molecules}

We now apply the symmetry-adapted algebraic approach to the vibrational 
excitations of two isotopes of ozone: $^{16}$O$_3$ and $^{18}$O$_3$.
In our analysis we include all vibrational excitation modes of ozone: 
the symmetric ($\nu_1$) and antisymmetric ($\nu_3$) stretching modes as 
well as the symmetric bending ($\nu_2$) mode.

In the electronic ground state the symmetric modes of ozone have the 
property of being almost pure stretching or bending modes \cite{chan}. 
Therefore we can simplify the Hamiltonian of (\ref{eq:halg3}) by 
neglecting the 
$\hat{\cal H}_{A_sA_b}$ term, which  couples the symmetric bending 
and stretching modes (as can be seen from  its harmonic limit in
(\ref{eq:hlim})).  
Furthermore, the strong anharmonic behavior of the antisymmetric $\nu_3$ 
mode suggests that higher order terms associated to this mode have to 
be taken into account in order to obtain results of spectroscopic quality. 
Here we propose to use the following quartic Hamiltonian to analyze the 
vibrational energies of ozone
\bea
\hat{\cal H}^{\prime} &=& \alpha_1 \, \hat{\cal H}_{A_s} 
+ \alpha_2 \, \hat{\cal H}_{A_b} + \alpha_3 \, \hat{\cal H}_{B_s} 
+ \beta_1 \, \hat{\cal V}_{1} + \beta_2 \, \hat{\cal V}_{2}
\nonumber\\ 
&& + \gamma_{33} \, (\hat{\cal H}_{B_s})^{2} + 
\gamma_{23} \, \hat{\cal H}_{A_b} \hat{\cal H}_{B_s} ~.
\label{eq:hamspec}
\eea
This Hamiltonian has seven adjustable parameters, plus two boson numbers, 
$N_1$ and $N_3$, whose values were fixed at $N_1=82$ and $N_3=68$.
A least-square fit to the vibrational spectrum of ozone by means of the
Hamiltonian (\ref{eq:hamspec}) yields a r.m.s. deviation for 
44 energies of 2.5 cm$^{-1}$. 
The r.m.s. deviation is defined in the usual way
\bea
\mbox{r.m.s.} &=& \sqrt{\sum_{i=1}^n (E_{exp}^i-E_{cal}^i)^2/(n-n_p)} ~,
\label{eq:rms}
\eea
where $n$ is the total number of energies in the fit and $n_p$ 
the number of parameters fitted. 
The values of the fitted parameters are presented in the second 
column of Table 1.
In Table 2 we compare the results of our calculation with the 
experimental energies, while in Table 3 we present the predicted energies 
for levels up to ten quanta which have not yet been observed. 
For a triatomic molecule with ${\cal{C}}_{2v}$ symmetry the usual 
notation for the vibrational labels corresponds to the spectroscopic 
(or normal) basis $(\nu_1,\nu_2,\nu_3)$ \cite{herzberg}.
In the tables we give the maximum component of the states in the normal 
and in the local basis. This allows  a 
selection of the appropriate labeling scheme in the presence of 
strong mixing. 
In general our theoretical assignments coincide with the experimental ones. 
At high  number of quanta, however, the strong mixing makes a unique
assignment  
difficult. Although for most of the levels the spectroscopic 
(or normal) labels are well determined, there are some levels with 
a strong local character. 
Our assignment of the stretching modes coincide with that of 
the analysis presented in \cite{kell1}.

In addition we have carried out the same fit in the harmonic limit 
($N_1$, $N_3 \rightarrow \infty$) of our 
model (see fourth column of Table 1). 
This calculation takes into account the harmonic counterparts of 
the operators in Hamiltonian (\ref{eq:hamspec}). Equation (\ref{eq:hlim}) 
shows that in this case the contribution of the $\beta_1$ and $\beta_2$ 
terms vanishes. The r.m.s. deviation increases from 2.5 to 73 cm$^{-1}$.
This shows that the strong anharmonicities that are present in the data 
can be accounted for in our approach by means of the terms proportional to 
$\beta_1$ and $\beta_2$, which do not have a counterpart in the harmonic 
limit, and only two anharmonicity constants, $\gamma_{23}$ and $\gamma_{33}$.

In previous calculations of ozone, both {\it ab initio} 
\cite{lee,Peterson} and algebraic \cite{chan,ben}, it was found 
important to include in addition to the harmonic frequencies $\omega_i$ 
and the anharmonicity constants $X_{ij}$, a Darling-Dennison coupling. 
We have repeated the previous calculation, 
but now including the nondiagonal contribution of a Darling-Dennison 
type coupling which is quartic in the generators \cite{ben}
\bea
\hat {\cal H}_{DD} &=&
\gamma \, \frac{1}{2N_1^2} \left[ (\hat J_-^{A_s})^2 (\hat J_+^{B_s})^2 + 
(\hat J_-^{B_s})^2 (\hat J_+^{A_s})^2 \right] ~, 
\eea
where
\bea
\hat J_{\pm}^{A_s} &=& \frac{1}{\sqrt{2}} 
( \hat J_{\pm,1} + \hat J_{\pm,2} ) ~, \nonumber\\
\hat J_{\pm}^{B_s} &=& \frac{1}{\sqrt{2}} 
( \hat J_{\pm,1} - \hat J_{\pm,2} ) ~.
\eea
In a fit to the same 44 levels (up to ten quanta) we find very little 
evidence  within our scheme 
 for the need of such a coupling {\it in addition} to the interaction 
terms already present in the Hamiltonian of (\ref{eq:hamspec}):
the value of the Darling-Dennison coupling strength found in the fit 
is very small, $\gamma= -0.074$ cm$^{-1}$ and the r.m.s. deviation is 
2.6 cm$^{-1}$, compared to 2.5 cm$^{-1}$ in the calculation of Table 2
which does not include this coupling.  
Although our basis is symmetry adapted, our calculations are carried 
out in terms of the local mode basis composed of independent anharmonic 
oscillators. In this sense our wave functions differ from the normal 
mode basis of \cite{ben} where the anharmonicity is introduced directly 
in the normal modes. 
As pointed out in \cite{xiao}, the DD coupling in the normal basis 
arises from the anharmonicity of the local modes. 
Thus, the purely anharmonic
interactions in the algebraic model  play a similar role to
that of (28). On the other hand, as we discuss below, in the
harmonic limit the $DD$ interaction is essential to achieve a good
quality fit to experiment. 

Previous calculations of ozone  have been limited 
to the stretching vibrations only \cite{chan,roo,ben,kell1}. 
The results presented in \cite{roo,ben,kell1} 
were obtained by fitting to the same set of 24 vibrational energies and 
show a standard deviation from experiment of 23.7, 9.1 and 7.9 cm$^{-1}$, 
respectively. The results of \cite{chan} were obtained by fitting 21 
vibrational energies with a standard deviation of 3.0 cm$^{-1}$. 
This is to be compared to a standard deviation of 2.4 cm$^{-1}$ 
in our calculations which  include both bending and stretching 
vibrations, and significantly more experimental data (a total of 44
vibrational  energies).  The  differences in the quality of the fits
may in part be due to uncertainties in the data.  For example, the
analysis in \cite{chan} and 
\cite{ben} was carried out with the same Hamiltonian,
but with a different data set leading to standard deviations of 3.0
and 9.1 cm$^{-1}$, respectively.   We note that the deviations
quoted in this paragraph are  standard deviations, which do not take
into account the number of parameters used in the fit.  Due to the
larger number of experimental energies, the difference between the
standard and the r.m.s. deviation in our calculation is much smaller
than in the calculations of \cite{chan,roo,ben,kell1}. 

On the other hand, the spectroscopic constants of ozone have been studied 
by many authors using {\it ab initio} methods. A recent calculation 
\cite{Peterson} shows  very good agreement  for the harmonic frequencies 
$\omega_i$, the anharmonicity constants $X_{ij}$, and the 
Darling-Dennison resonance constant $\gamma$, as compared to the values 
extracted from experiment (up to four quanta) \cite{barbe}.
The standard deviation between the vibrational spectrum generated 
from these constants and 17 observed vibrational energies (up to three 
quanta  and one state corresponding to four quanta) is 8.3 cm$^{-1}$ 
(see Table 5  of \cite{Peterson}). When we repeat this calculation and 
extend it to the vibrational energies corresponding to the 
44 experimental values of Table 2, however, the standard deviation 
increases to 53.2 cm$^{-1}$.  In a calculation 
in which the same parameters  $\omega_i$, $X_{ij}$  
and $\gamma$ are fitted to the 44 experimental 
vibrational energies of ozone, we find a  standard deviation 
of 2.3 cm$^{-1}$, which is comparable to the fit presented in Table
2.  The latter calculation corresponds in our model to the harmonic
limit, which includes the DD interaction.  

Finally, we present the results of a calculation of the vibrational 
energies of $^{18}$O$_3$. For the two isotopes $^{16}$O$_3$ and 
$^{18}$O$_3$ the boson numbers scale with the square root of the mass, 
the harmonic frequencies with the inverse square root of the mass and 
the anharmonicity constants with the inverse mass. Therefore we take 
\bea
^{18}N_i+1 &=& \left( \frac{^{18}\mu}{^{16}\mu} \right)^{1/2} \, 
(^{16}N_i+1) ~,
\nonumber\\
^{18}\alpha_i &=& \left( \frac{^{16}\mu}{^{18}\mu} \right)^{1/2} \, 
^{16}\alpha_i ~,
\nonumber\\
^{18}\gamma_{ij} &=& \left( \frac{^{16}\mu}{^{18}\mu} \right) \, 
^{16}\gamma_{ij} ~. 
\label{eq:scale}
\eea
The remaining two parameters $\beta_1$ and $\beta_2$ are fitted to 
the 15 vibrational energies of $^{18}$O$_3$ with a r.m.s. deviation 
of 2.3 cm$^{-1}$ (see second column of Table 4). 
This is of the same accuracy as our fit to 
$^{16}$O$_3$ which has a r.m.s. deviation of 2.5 cm$^{-1}$. 
In a second calculation we keep the boson numbers fixed at $N_1=87$
and $N_3=72$ and fit the remaining seven parameters to the data. 
The result is a fit with a r.m.s. deviation of 1.0 cm$^{-1}$ 
(see third column of Table 4).
The parameters for $^{18}$O$_3$ are shown in Table 4, and the calculated 
vibrational energies are compared to the experimental data in Table 5. 
In Table 6 we present the predicted energies for levels 
up to four quanta which have not yet been observed. The two calculations 
presented in Tables 5 and 6 
correspond to the parameter sets given in the second and third column 
of Table 4. The maximum component of the states in the local and normal 
basis in Tables 5 and 6 correspond to the first calculation.

\section{Summary and conclusions}

We have presented a study of the full vibrational spectrum of ozone, 
including both stretching and bending modes, 
by means of a symmetry adapted  algebraic model. A least-square energy fit 
to 44 levels of $^{16}$O$_3$ with seven parameters and two 
boson numbers yields a r.m.s. deviation of 2.5 cm$^{-1}$. 
We also  calculated the vibrational energies of $^{18}$O$_3$ by 
appropriately scaling the boson numbers, the harmonic frequencies and 
the anharmonicity constants with the masses. The two remaining parameters 
were determined in a least-square fit to 15 energies with a r.m.s. 
deviation of 2.3 cm$^{-1}$. The quality of these fits show that the model 
is able to give an accurate description of the 
vibrational energies of the two ozone isotopes. Since the
values of the fitted parameters change very little when 
states with higher number of quanta are added successively,  
the model may provide a reliable alternative to integro-differential 
methods as an easy-to-handle tool with 
considerable predictive power, which can be particularly useful when
no {\it ab initio} calculations are available.  

An interesting point concerns the role of the Darling-Dennison 
coupling between the stretching modes.  In our analysis of the experimental 
vibrational energies up to ten quanta we find no need to include this 
coupling explicitly {\it in addition} to the interaction terms
already present  
in the algebraic Hamiltonian. 
The strong anharmonicities that are present in the data 
can be accounted for in our approach by means of the terms proportional to 
$\beta_1$ and $\beta_2$, which do not have a counterpart in the harmonic 
limit, together with the  anharmonicity constants, $\gamma_{23}$ and
$\gamma_{33}$. 
In a calculation in the harmonic limit, all six anharmonicities as well 
as a Darling-Dennison coupling have to be included to achieve a fit of 
comparable quality.

The main difference  between the present model and 
  previous algebraic models  is the use 
of symmetry-adapted interactions, which in the harmonic limit have a
direct connection with the conventional configuration space
interactions.  It also provides a systematic procedure to 
incorporate higher order terms which have a selective effect on the  
various modes. From a numerical point of view the new operators 
significantly improve the convergence of the fits.  We  point
out, however, that due to the rather simple symmetry characteristics
of the ozone molecule, these advantages are in this case not as
patent as they are for more complex molecules, such as Be$_4$ 
\cite{letter} and CH$_4$ \cite{ch4}.  

In addition to the energies, the algebraic model provides wave 
functions, which can be used to compute other observables, such 
as transition intensities. The algebraic transition operators can be 
constructed using the connection between algebraic and 
configuration space operators, which was discussed in \cite{letter} 
with respect to the Hamiltonian operator. An analysis of the infrared 
band intensities for the ozone molecule  is in progress.  

\section*{Acknowledgements}

This work was supported in part by the European Community under 
contract nr CI1*-CT94-0072, DGAPA-UNAM, M\'exico, under project IN105194, 
CONACyT-M\'exico under project 400340-5-3401E and Spanish DGICyT under 
project PB92-0663.

\clearpage

\section*{Table  Captions}

\noindent Table 1.- 
Parameters and boson numbers used in the calculation of 
$^{16}$O$_3$ energy levels.
Parameters and r.m.s. are given in cm$^{-1}$. 

\bigskip

\noindent Table 2.-  
Fit to $^{16}$O$_3$ vibrational spectrum. 
Here $\Delta E_{cal}=E_{exp}-E_{cal}$.
The experimental data are taken from \cite{atlas,exp}.  
Energies and r.m.s. are given in cm$^{-1}$. 

\bigskip

\noindent  Table 3.- 
Predicted $^{16}$O$_3$ vibrational energy levels up to  ten quanta.
Energies are given in cm$^{-1}$.  

\bigskip

\noindent Table 4.- 
Parameters and boson numbers used in the calculation of 
$^{18}$O$_3$ energy levels. See text for explanation of the two sets 
of parameters.  Parameters and r.m.s. are given in cm$^{-1}$.

\bigskip

\noindent  Table 5.- 
Fit to $^{18}$O$_3$ vibrational spectrum. 
Here $\Delta E_{cal}=E_{exp}-E_{cal}$. The calculations 
correspond to the two parameter sets of Table 4.
The experimental data are taken from \cite{barbe}. 
Energies and r.m.s. are given in cm$^{-1}$.

\bigskip

\noindent  Table 6.- 
Predicted $^{18}$O$_3$ vibrational energy levels 
up to four quanta. The calculations 
correspond to the two parameter sets of Table 4. 
Energies are given in cm$^{-1}$.

\clearpage

\section*{Figure Captions}

\noindent Figure 1.- 
Geometrical structure of ozone molecule showing the 
assignment of the $U(2)$ algebraic structure.

\bigskip

\noindent Figure 2.- 
Eigenvalues of the operator $\hat {\cal H}_{A_s}$ 
as a function of the number of bosons $N_1$ for 
(a) two quanta and (b) three quanta. 
The numbers in parentheses denote normal basis labels 
$(\nu_1,\nu_2,\nu_3)$.

\bigskip

\noindent Figure 3.- 
Eigenvalues of the operator $\hat {\cal H}_{A_b}$ 
as a function of the number of bosons $N_3$ for 
(a) two quanta and (b) three quanta.
The numbers in parentheses denote normal basis labels 
$(\nu_1,\nu_2,\nu_3)$.

\bigskip

\noindent Figure 4.- 
Eigenvalues of the operator $\hat {\cal V}_{1}$ 
as a function of the number of bosons $N_1$ for 
(a) two quanta and (b) three quanta.
The numbers in parentheses denote local basis labels 
$(\nu_1,\nu_2,\nu_3)$.

\clearpage

\begin{table}
\centering
\caption[]
{Parameters and boson numbers used in the calculation of 
$^{16}$O$_3$ energy levels.
Parameters and r.m.s. are given in cm$^{-1}$.}
\label{par16}
\vspace{10pt}
\begin{tabular}{|c|r|c|r|}          
\hline
\hline
\multicolumn{2}{|c|}{Algebraic Model} &
\multicolumn{2}{|c|}{Harmonic Limit}  \\
\hline
$N_1$         &   $82$   & $N_1$      & $\infty$ \\ 
$N_3$         &   $68$   & $N_3$      & $\infty$ \\
$\beta_1$     & $1884.0$ & $-$        &      $-$ \\
$\beta_2$     & $-316.5$ & $-$        &      $-$ \\                   
$\alpha_1$    & $1099.8$ & $\omega_1$ &   $1071$ \\
$\alpha_2$    &  $700.0$ & $\omega_2$ &    $702$ \\
$\alpha_3$    & $1046.6$ & $\omega_3$ &   $1007$ \\
$\gamma_{23}$ &   $-9.5$ & $X_{23}$   &      $3$ \\
$\gamma_{33}$ &   $-5.7$ & $X_{33}$   &     $-8$ \\
\hline
r.m.s. & $2.5$ & r.m.s. & 73 \\
\hline
\hline
\end{tabular}
\end{table}

\clearpage
\begin{table}
\centering
\caption[]{Fit to $^{16}$O$_3$ vibrational spectrum. 
Here $\Delta E_{cal}=E_{exp}-E_{cal}$.
The experimental data are taken from [1,2].
Energies and r.m.s. are given in cm$^{-1}$.}
\label{fit16}
\vspace{10pt}
\begin{tabular}{|c|c|r|r|}          
\hline
\hline
Local Basis & Normal Basis & $E_{exp}$ & $\Delta E_{cal}$ \\
$(v_1,v_2,v_3)$ & $(\nu_1,\nu_2,\nu_3)$ & & \\
\hline        
   ( 0 0+  1)  &   ( 0 1 0)   &  700.9  &  1.0  \\
   ( 0 1-- 0)  &   ( 0 0 1)   & 1042.1  &--2.3  \\
   ( 1 0+  0)  &   ( 1 0 0)   & 1103.1  &--0.2  \\
   ( 0 0+  2)  &   ( 0 2 0)   & 1399.3  &--0.3  \\
   ( 1 0-- 1)  &   ( 0 1 1)   & 1726.5  &  0.0  \\
   ( 0 1+  1)  &   ( 1 1 0)   & 1796.3  &  1.6  \\
   ( 0 2+  0)  &   ( 0 0 2)   & 2057.9  &--3.8  \\
   ( 2 0-- 0)  &   ( 1 0 1)   & 2110.8  &  2.0  \\
   ( 1 1+  0)  &   ( 2 0 0)   & 2201.2  &--1.6  \\
   ( 0 1-- 2)  &   ( 0 2 1)   & 2407.9  &--0.8  \\
   ( 1 0+  2)  &   ( 1 2 0)   & 2486.6  &  0.6  \\ 
   ( 2 0+  1)  &   ( 0 1 2)   & 2726.1  &--0.7  \\
   ( 0 2-- 1)  &   ( 1 1 1)   & 2785.2  &  2.8  \\
   ( 1 1+  1)  &   ( 2 1 0)   & 2886.2  &  1.2  \\
   ( 0 3-- 0)  &   ( 0 0 3)   & 3046.1  &--2.9  \\
   ( 3 0+  0)  &   ( 1 0 2)   & 3083.7  &  3.0  \\
   ( 1 2-- 0)  &   ( 2 0 1)   & 3186.4  &  1.0  \\
   ( 1 2+  0)  &   ( 3 0 0)   & 3290.0  &--2.2  \\
   ( 2 0+  2)  &   ( 0 2 2)   & 3395.1  &  2.9  \\
   ( 0 2-- 2)  &   ( 1 2 1)   & 3455.8  &--0.3  \\
   ( 1 1+  2)  &   ( 2 2 0)   & 3563.5  &--3.7  \\
   ( 3 0-- 1)  &   ( 0 1 3)   & 3698.3  &--0.1  \\
   ( 3 0+  1)  &   ( 1 1 2)   & 3739.4  &  1.8  \\
   ( 1 2-- 1)  &   ( 2 1 1)   & 3849.9  &  2.0  \\
   ( 2 1+  1)  &   ( 3 1 0)   & 3966.6  &  0.9  \\
   ( 4 0+  0)  &   ( 0 0 4)   & 4001.4  &  1.1  \\
   ( 0 4-- 0)  &   ( 1 0 3)   & 4021.8  &  4.3  \\
   ( 2 2+  0)  &   ( 2 0 2)   & 4139.0  &--3.9  \\
   ( 3 0-- 2)  &   ( 0 2 3)   & 4346.7  &--1.3  \\
   ( 2 2+  0)  &   ( 4 0 0)   & 4371.0  &--1.8  \\
   ( 3 0+  2)  &   ( 1 2 2)   & 4392.5  &--2.3  \\
   ( 5 0+  0)  &   ( 1 0 4)   & 4922.0  &  5.1  \\
   ( 4 1+  0)  &   ( 3 0 2)   & 5170.0  &--0.8  \\
   ( 2 3+  0)  &   ( 5 0 0)   & 5443.0  &--0.6  \\
   ( 6 0+  0)  &   ( 2 0 4)   & 5767.0  &--7.1  \\
   ( 2 4+  0)  &   ( 0 0 6)   & 5997.0  &  1.7  \\
   ( 3 3+  0)  &   ( 4 0 2)   & 6204.0  &--1.6  \\
   ( 3 3+  0)  &   ( 6 0 0)   & 6506.0  &  1.2  \\
   ( 6 1+  0)  &   ( 1 0 6)   & 6927.0  &  2.5  \\
   ( 2 5+  0)  &   ( 5 0 2)   & 7227.0  &--0.3  \\
   ( 4 3+  0)  &   ( 7 0 0)   & 7555.0  &--1.2  \\
   ( 4 4+  0)  &   ( 8 0 0)   & 8598.0  &  0.2  \\
   ( 4 5+  0)  &   ( 9 0 0)   & 9632.0  &  2.6  \\
   ( 5 5+  0)  &   (10 0 0)   &10650.0  &--1.1  \\   
\hline					 
 r.m.s.     &       &         &  2.5  \\
\hline
\hline
\end{tabular}
\end{table}

\clearpage
\footnotesize
\begin{table}
\centering
\caption[]{Predicted $^{16}$O$_3$ vibrational energy levels up to 
ten quanta. Energies are given in cm$^{-1}$.}
\label{vib16}
\vspace{10pt}
\begin{tabular}{|c|c|r|c|c|r|}          
\hline
\hline
Local Basis & Normal Basis & $E_{cal}$ & 
Local Basis & Normal Basis & $E_{cal}$ \\  
$(v_1,v_2,v_3)$ & $(\nu_1,\nu_2,\nu_3)$ & & 
$(v_1,v_2,v_3)$ & $(\nu_1,\nu_2,\nu_3)$ & \\
\hline        
  ( 0 0+ 3)  & ( 0 3 0) &  2099.3  & ( 2 2+ 3)   & ( 2 3 2) & 6074.2  \\  
  ( 0 0+ 4)  & ( 0 4 0) &  2798.7  & ( 3 2+ 1)   & ( 5 1 0) & 6099.5  \\  
  ( 1 0-- 3) & ( 0 3 1) &  3091.0  & ( 0 2-- 6)  & ( 1 6 1) & 6152.5  \\ 
  ( 0 1+ 3)  & ( 1 3 0) &  3177.1  & ( 5 0-- 2)  & ( 2 2 3) & 6155.9  \\
  ( 0 0+ 5)  & ( 0 5 0) &  3498.0  & ( 0 5+ 2)   & ( 1 2 4) & 6170.1  \\
  ( 1 0-- 4) & ( 0 4 1) &  3773.5  & ( 1 3-- 3)  & ( 3 3 1) & 6210.4  \\
  ( 1 0+ 4)  & ( 1 4 0) &  3868.1  & ( 0 0+ 9)   & ( 0 9 0) & 6293.8  \\
  ( 1 1+ 3)  & ( 0 3 2) &  4057.9  & ( 1 1+ 6)   & ( 2 6 0) & 6295.6  \\
  ( 2 0-- 3) & ( 1 3 1) &  4130.0  & ( 1 2-- 5)  & ( 0 5 3) & 6298.7  \\
  ( 0 0+ 6)  & ( 0 6 0) &  4197.2  & ( 2 3-- 2)  & ( 0 2 5) & 6326.6  \\
  ( 1 3-- 0) & ( 3 0 1) &  4247.5  & ( 4 2-- 0)  & ( 5 0 1) & 6352.4  \\
  ( 1 1+ 3)  & ( 2 3 0) &  4249.5  & ( 2 2+ 3)   & ( 4 3 0) & 6366.8  \\
  ( 0 1-- 5) & ( 0 5 1) &  4456.2  & ( 3 0+ 5)   & ( 1 5 2) & 6368.0  \\
  ( 1 2-- 2) & ( 2 2 1) &  4511.1  & ( 0 6+ 1)   & ( 2 1 4) & 6385.3  \\
  ( 1 0+ 5)  & ( 1 5 0) &  4559.0  & ( 6 0-- 1)  & ( 3 1 3) & 6389.3  \\
  ( 0 4+ 1)  & ( 0 1 4) &  4636.0  & ( 1 4+ 2)   & ( 3 2 2) & 6441.7  \\
  ( 2 1+ 2)  & ( 3 2 0) &  4639.2  & ( 3 0-- 5)  & ( 2 5 1) & 6503.6  \\
  ( 4 0-- 1) & ( 1 1 3) &  4658.7  & ( 1 0-- 8)  & ( 0 8 1) & 6504.8  \\
  ( 1 1+ 4)  & ( 0 4 2) &  4723.8  & ( 4 0+ 4)   & ( 0 4 4) & 6542.6  \\
  ( 2 2+ 1)  & ( 2 1 2) &  4785.3  & ( 0 4-- 4)  & ( 1 4 3) & 6584.3  \\
  ( 2 0-- 4) & ( 1 4 1) &  4804.0  & ( 4 1-- 2)  & ( 4 2 1) & 6593.8  \\
  ( 0 0+ 7)  & ( 0 7 0) &  4896.2  & ( 7 0-- 0)  & ( 4 0 3) & 6595.1  \\
  ( 3 1-- 1) & ( 3 1 1) &  4901.3  & ( 0 7+ 0)   & ( 3 0 4) & 6596.0  \\
  ( 0 5-- 0) & ( 2 0 3) &  4909.4  & ( 3 3+ 1)   & ( 0 1 6) & 6599.7  \\
  ( 1 1+ 4)  & ( 2 4 0) &  4931.6  & ( 0 1+ 8)   & ( 1 8 0) & 6630.6  \\
  ( 3 0-- 3) & ( 0 3 3) &  4997.8  & ( 2 1+ 5)   & ( 3 5 0) & 6659.5  \\
  ( 2 2+ 1)  & ( 4 1 0) &  5037.4  & ( 1 5-- 1)  & ( 1 1 5) & 6680.6  \\
  ( 3 0+ 3)  & ( 1 3 2) &  5052.2  & ( 2 2+ 4)   & ( 2 4 2) & 6720.2  \\
  ( 2 3-- 0) & ( 0 0 5) &  5080.0  & ( 1 1+ 7)   & ( 0 7 2) & 6723.9  \\
  ( 0 1-- 6) & ( 0 6 1) &  5138.9  & ( 3 2+ 2)   & ( 5 2 0) & 6755.4  \\
  ( 2 1-- 3) & ( 2 3 1) &  5174.9  & ( 0 5-- 3)  & ( 2 3 3) & 6778.8  \\
  ( 1 0+ 6)  & ( 1 6 0) &  5249.7  & ( 5 0+ 3)   & ( 1 3 4) & 6797.2  \\
  ( 4 0+ 2)  & ( 0 2 4) &  5271.5  & ( 2 0-- 7)  & ( 1 7 1) & 6826.9  \\
  ( 4 0-- 2) & ( 1 2 3) &  5300.2  & ( 3 3+ 1)   & ( 4 1 2) & 6829.7  \\
  ( 1 4-- 0) & ( 4 0 1) &  5304.8  & ( 3 1-- 4)  & ( 3 4 1) & 6865.6  \\
  ( 2 1+ 3)  & ( 3 3 0) &  5312.7  & ( 6 1-- 0)  & ( 0 0 7) & 6882.0  \\
  ( 1 1+ 5)  & ( 0 5 2) &  5390.2  & ( 2 1-- 6)  & ( 0 6 3) & 6949.8  \\
  ( 2 2+ 2)  & ( 2 2 2) &  5429.2  & ( 3 2-- 3)  & ( 0 3 5) & 6952.3  \\
  ( 2 0-- 5) & ( 1 5 1) &  5478.2  & ( 1 1+ 7)   & ( 2 7 0) & 6977.5  \\
  ( 5 0-- 1) & ( 2 1 3) &  5532.8  & ( 4 2-- 1)  & ( 5 1 1) & 6987.5  \\
  ( 5 0+ 1)  & ( 1 1 4) &  5543.4  & ( 0 0+10)   & ( 010 0) & 6992.4  \\
  ( 1 3-- 2) & ( 3 2 1) &  5555.6  & ( 0 6+ 2)   & ( 2 2 4) & 6996.3  \\
  ( 0 0+ 8)  & ( 0 8 0) &  5595.1  & ( 0 6-- 2)  & ( 1 2 5) & 7002.1  \\
  ( 1 1+ 5)  & ( 2 5 0) &  5613.7  & ( 0 3+ 6)   & ( 1 6 2) & 7026.4  \\
  ( 0 3-- 4) & ( 0 4 3) &  5648.0  & ( 2 2+ 4)   & ( 4 4 0) & 7031.5  \\
  ( 2 2+ 2)  & ( 4 2 0) &  5702.1  & ( 4 1+ 3)   & ( 3 3 2) & 7078.4  \\
  ( 2 3-- 1) & ( 0 1 5) &  5702.4  & ( 4 3-- 0)  & ( 2 0 5) & 7082.0  \\
  ( 3 0+ 4)  & ( 1 4 2) &  5710.0  & ( 3 3+ 1)   & ( 6 1 0) & 7151.9  \\
  ( 0 6-- 0) & ( 3 0 3) &  5776.8  & ( 0 3-- 6)  & ( 2 6 1) & 7168.5  \\
  ( 4 1+ 1)  & ( 3 1 2) &  5805.8  & ( 0 4+ 5)   & ( 0 5 4) & 7178.7  \\
  ( 1 0-- 7) & ( 0 7 1) &  5821.8  & ( 0 1-- 9)  & ( 0 9 1) & 7188.0  \\
  ( 2 1-- 4) & ( 2 4 1) &  5839.1  & ( 0 7-- 1)  & ( 2 1 5) & 7193.5  \\
  ( 0 4+ 3)  & ( 0 3 4) &  5907.0  & ( 7 0+ 1)   & ( 3 1 4) & 7194.9  \\
  ( 0 1+ 7)  & ( 1 7 0) &  5940.2  & ( 3 3+ 2)   & ( 0 2 6) & 7205.4  \\
  ( 4 0-- 3) & ( 1 3 3) &  5942.1  & ( 4 0-- 5)  & ( 1 5 3) & 7227.0  \\
  ( 1 4-- 1) & ( 4 1 1) &  5948.9  & ( 1 4-- 3)  & ( 4 3 1) & 7239.3  \\
  ( 2 1+ 4)  & ( 3 4 0) &  5986.1  & ( 5 1-- 2)  & ( 3 2 3) & 7298.7  \\
  ( 1 1+ 6)  & ( 0 6 2) &  6056.9  &  ( 1 0+ 9)  & ( 1 9 0) & 7320.9  \\
  ( 5 1-- 0) & ( 1 0 5) &  6063.4  &  ( 2 1+ 6)  & ( 3 6 0) & 7332.8  \\
\hline			 
\hline
\end{tabular}
\end{table}

\clearpage
\footnotesize
\begin{table}
\centering
\addtocounter{table}{-1}
\caption[]{Continued}
\vspace{10pt}
\begin{tabular}{|c|c|r|c|c|r|}          
\hline
\hline
Local Basis & Normal Basis & $E_{cal}$ &
Local Basis & Normal Basis & $E_{cal}$ \\  
$(v_1,v_2,v_3)$ & $(\nu_1,\nu_2,\nu_3)$ & & 
$(v_1,v_2,v_3)$ & $(\nu_1,\nu_2,\nu_3)$ & \\
\hline                                  
 ( 2 2+ 5)  & ( 2 5 2) & 7367.1 &  ( 7 0-- 3) & ( 2 3 5) &  8390.9  \\
 ( 8 0+ 0)  & ( 4 0 4) & 7373.4 &  ( 0 7+ 3) & ( 3 3 4)  &  8393.7  \\  
 ( 0 8-- 0) & ( 3 0 5) & 7373.6 &  ( 3 3+ 4) & ( 0 4 6)  &  8421.3  \\
 ( 1 1+ 8)  & ( 0 8 2) & 7391.4 &  ( 3 5-- 0) & ( 7 0 1) &  8421.8  \\
 ( 2 5-- 0) & ( 6 0 1) & 7391.7 &  ( 3 3+ 3) & ( 6 3 0)  &  8446.4  \\
 ( 0 5-- 4) & ( 0 4 5) & 7401.6 &  ( 2 5+ 2) & ( 5 2 2)  &  8458.9  \\
 ( 2 3+ 3)  & ( 5 3 0) & 7411.4 &  ( 4 4+ 1) & ( 0 1 8)  &  8522.2  \\
 ( 0 5+ 4)  & ( 1 4 4) & 7424.8 &  ( 1 4-- 5)& ( 4 5 1)  &  8531.8  \\
 ( 1 5+ 2)  & ( 4 2 2) & 7455.2 &  ( 8 1-- 0) & ( 2 0 7) &  8533.2  \\
 ( 5 2-- 1) & ( 0 1 7) & 7471.3 &  ( 1 5-- 4) & ( 3 4 3) &  8538.0  \\
 ( 2 0-- 8) & ( 1 8 1) & 7501.4 &  ( 8 1+ 0) & ( 1 0 8)  &  8542.0  \\
 ( 3 1-- 5) & ( 3 5 1) & 7521.2 &  ( 0 8+ 2) & ( 2 2 6)  &  8544.4  \\
 ( 1 6+ 1)  & ( 1 1 6) & 7524.9 &  ( 8 0-- 2) & ( 3 2 5) &  8545.0  \\
 ( 2 3-- 4) & ( 2 4 3) & 7579.6 &  ( 2 5-- 2) & ( 6 2 1) &  8644.1  \\
 ( 2 1-- 7) & ( 0 7 3) & 7601.5 &  ( 2 5-- 3) & ( 0 3 7) &  8652.0  \\
 ( 0 6+ 3)  & ( 2 3 4) & 7607.3 &  ( 2 6-- 1) & ( 3 1 5) &  8668.1  \\
 ( 6 0-- 3) & ( 3 3 3) & 7615.2 &  ( 9 0-- 1) & ( 4 1 5) &  8681.5  \\
 ( 2 4-- 2) & ( 5 2 1) & 7623.2 &  ( 0 9+ 1) & ( 3 1 6)  &  8681.6  \\
 ( 1 1+ 8)  & ( 2 8 0) & 7659.2 &  ( 1 5+ 4) & ( 4 4 2)  &  8709.5  \\
 ( 3 0+ 7)  & ( 1 7 2) & 7685.2 &  ( 3 2+ 5) & ( 5 5 0)  &  8723.4  \\
 ( 1 6-- 1) & ( 4 1 3) & 7685.4 &  ( 6 1+ 3) & ( 1 3 6)  &  8728.7  \\
 ( 2 2+ 5)  & ( 4 5 0) & 7696.2 &  ( 5 4-- 0) & ( 0 0 9) &  8783.4  \\
 ( 4 1+ 4)  & ( 3 4 2) & 7715.8 &  ( 010+ 0) & ( 4 0 6)  &  8803.5  \\
 ( 7 1+ 0)  & ( 0 0 8) & 7730.3 &  (10 0-- 0) & ( 5 0 5) &  8803.5  \\
 ( 1 7-- 0) & ( 1 0 7) & 7751.7 &  ( 3 4+ 2) & ( 7 2 0)  &  8833.1  \\
 ( 0 7-- 2) & ( 2 2 5) & 7792.1 &  ( 6 2+ 1) & ( 6 1 2)  &  8849.1  \\
 ( 7 0+ 2)  & ( 3 2 4) & 7794.1 &  ( 1 7+ 2) & ( 0 2 8)  &  8884.9  \\
 ( 3 3+ 2)  & ( 6 2 0) & 7799.1 &  ( 2 7+ 0) & ( 1 0 8)  &  8888.6  \\
 ( 3 3+ 3)  & ( 0 3 6) & 7812.6 &  ( 4 2-- 4) & ( 5 4 1) &  8896.6  \\
 ( 4 0+ 6)  & ( 0 6 4) & 7815.2 &  ( 1 6-- 3) & ( 2 3 5) &  8898.3  \\
 ( 3 0-- 7) & ( 2 7 1) & 7833.7 &  ( 7 1-- 2) & ( 1 2 7) &  8922.3  \\
 ( 5 2+ 1)  & ( 5 1 2) & 7842.5 &  ( 3 5-- 1) & ( 7 1 1) &  9038.5  \\
 ( 0 4-- 6) & ( 1 6 3) & 7870.1 &  ( 2 7-- 0) & ( 4 0 5) &  9066.2  \\
 ( 4 1-- 4) & ( 4 4 1) & 7885.3 &  ( 5 2+ 3) & ( 5 3 2)  &  9076.4  \\
 ( 1 5-- 3) & ( 3 3 3) & 7917.9 &  ( 3 3+ 4) & ( 6 4 0)  &  9093.7  \\
 ( 4 4+ 0)  & ( 2 0 6) & 7940.4 &  ( 1 8-- 1) & ( 0 1 9) &  9099.9  \\
 ( 8 0+ 1)  & ( 2 1 6) & 7958.7 &  ( 4 4+ 2) & ( 2 2 6) &   9106.8  \\ 
 ( 0 8-- 1) & ( 3 1 5) & 7959.1 &  ( 8 1+ 1) & ( 1 1 8) &   9112.5  \\ 
 ( 1 2+ 7)  & ( 3 7 0) & 8006.0 &  ( 4 4+ 1) & ( 8 1 0) &   9227.4  \\ 
 ( 2 2+ 6)  & ( 2 6 2) & 8014.9 &  ( 6 3+ 0) & ( 7 0 2) &   9250.5  \\ 
 ( 2 5-- 1) & ( 6 1 1) & 8017.5 &  ( 7 1-- 2) & ( 3 2 5) &  9265.6  \\ 
 ( 5 0-- 5) & ( 0 5 5) & 8024.6 &  ( 2 5-- 3) & ( 6 3 1) &  9271.4  \\ 
 ( 0 5+ 5)  & ( 1 5 4) & 8052.8 &  ( 1 9+ 0) & ( 2 0 8) &   9290.0  \\ 
 ( 5 2-- 2) & ( 0 2 7) & 8061.2 &  ( 9 1-- 0) & ( 1 0 9) &  9293.1  \\ 
 ( 2 3+ 4)  & ( 5 4 0) & 8067.4 &  ( 4 5-- 1) & ( 0 1 9) &  9344.6  \\ 
 ( 6 2-- 0) & ( 3 0 5) & 8072.1 &  ( 6 3-- 0) & ( 8 0 1) &  9442.7  \\ 
 ( 1 5+ 3)  & ( 4 3 2) & 8081.8 &  ( 6 2+ 2) & ( 6 2 2) &   9455.9  \\ 
 ( 0 9-- 0) & ( 4 0 5) & 8109.4 &  ( 6 3+ 1) & ( 3 1 6) &   9465.9  \\ 
 ( 9 0+ 0)  & ( 3 0 6) & 8109.5 &  ( 4 3+ 3) & ( 7 3 0) &   9471.6  \\ 
 ( 1 6+ 2)  & ( 1 2 6) & 8126.3 &  ( 3 7+ 0) & ( 0 010) &   9604.9  \\ 
 ( 1 3-- 6) & ( 3 6 1) & 8177.1 &  ( 7 2-- 1) & ( 4 1 5) &  9651.5  \\ 
 ( 3 4+ 1)  & ( 7 1 0) & 8194.6 &  ( 3 5-- 2) & ( 7 2 1) &  9656.0  \\ 
 ( 2 3-- 5) & ( 2 5 3) & 8208.3 &  ( 7 3-- 0) & ( 1 0 9) &  9676.2  \\ 
 ( 6 0+ 4)  & ( 2 4 4) & 8218.4 &  ( 2 7+ 1) & ( 7 1 2) &   9846.8  \\ 
 ( 0 6-- 4) & ( 1 4 5) & 8228.8 &  ( 4 4+ 2) & ( 8 2 0) &   9857.2  \\ 
 ( 6 2+ 0)  & ( 6 0 2) & 8243.6 &  ( 8 2+ 0) & ( 2 0 8) &   9865.4  \\ 
 ( 4 2-- 3) & ( 5 3 1) & 8259.6 &  ( 2 8-- 0) & ( 9 0 1) & 10048.0  \\ 
 ( 1 6-- 2) & ( 4 2 3) & 8290.9 &  ( 6 3-- 1) & ( 8 1 1) & 10050.3  \\ 
 ( 1 7+ 1)  & ( 0 1 8) & 8307.6 &  ( 7 3+ 0) & ( 6 0 4) &  10249.3  \\ 
 ( 1 7-- 1) & ( 1 1 7) & 8336.6 &  ( 4 5+ 1) & ( 9 1 0) &  10250.3  \\ 
 ( 4 1+ 5)  & ( 3 5 2) & 8354.0 &  ( 4 6-- 0) & ( 9 0 1) & 10454.2  \\ 
 ( 2 2+ 6)  & ( 4 6 0) & 8360.8 &             &          &          \\
\hline 
\hline                                                           
\end{tabular}
\end{table}

\clearpage
\normalsize
\begin{table}
\centering
\caption[]{Parameters and boson numbers used in the calculation of 
$^{18}$O$_3$ energy levels. See text for explanation of the two sets 
of parameters.
Parameters and r.m.s. are given in cm$^{-1}$.}
\label{par18}
\vspace{10pt}
\begin{tabular}{|c|r|r|}          
\hline
\hline
$N_1$         & $87$     & $87$     \\ 
$N_3$         & $72$     & $72$     \\
$\beta_1$     & $1731.3$ & $1675.8$ \\
$\beta_2$     & $-340.4$ & $-292.4$ \\                   
$\alpha_1$    & $1037.9$ & $1039.4$ \\
$\alpha_2$    &  $660.0$ &  $661.1$ \\
$\alpha_3$    &  $987.1$ &  $987.5$ \\
$\gamma_{23}$ &   $-8.4$ &   $-7.5$ \\
$\gamma_{33}$ &   $-5.1$ &   $-5.4$ \\
\hline
r.m.s.        &    $2.3$ &    $1.0$ \\
\hline
\hline
\end{tabular}
\end{table}

\clearpage
\begin{table}
\centering
\caption[]{Fit to $^{18}$O$_3$ vibrational spectrum. 
Here $\Delta E_{cal}=E_{exp}-E_{cal}$. The calculations 
correspond to the two parameter sets of Table \ref{par18}.
The experimental data are taken from [4]. 
Energies and r.m.s. are given in cm$^{-1}$.}
\label{fit18}
\vspace{10pt}
\begin{tabular}{|c|c|r|r|r|}          
\hline
\hline
Local Basis & Normal Basis & $E_{exp}$ & $\Delta E_{cal}$ 
& $\Delta E_{cal}$ \\  
$(v_1,v_2,v_3)$ & $(\nu_1,\nu_2,\nu_3)$ & & & \\
\hline        
   ( 0 0+  1)  &  ( 0 1 0)  &  661.7  &   0.5 &   0.9 \\
   ( 0 1-- 0)  &  ( 0 0 1)  &  984.6  & --1.0 & --0.5 \\
   ( 1 0+  0)  &  ( 1 0 0)  & 1041.9  &   0.4 & --0.5 \\
   ( 0 1-- 1)  &  ( 0 1 1)  & 1631.2  &   1.3 &   0.0 \\
   ( 1 0+  1)  &  ( 1 1 0)  & 1695.9  &   1.8 &   0.0 \\
   ( 0 2+  0)  &  ( 0 0 2)  & 1945.4  & --3.1 & --1.3 \\
   ( 2 0-- 0)  &  ( 1 0 1)  & 1995.1  &   0.7 & --0.0 \\
   ( 1 1+  0)  &  ( 2 0 0)  & 2079.4  &   0.3 & --0.7 \\ 
   ( 2 0+  1)  &  ( 0 1 2)  & 2579.5  &   3.0 &   1.0 \\
   ( 0 2-- 1)  &  ( 1 1 1)  & 2634.3  &   4.2 &   0.5 \\
   ( 3 0-- 0)  &  ( 0 0 3)  & 2883.2  & --3.5 &   0.0 \\
   ( 1 2-- 0)  &  ( 2 0 1)  & 3012.6  &   1.4 &   1.6 \\
   ( 0 2-- 2)  &  ( 1 2 1)  & 3271.0  &   2.5 & --1.2 \\
   ( 3 0-- 1)  &  ( 0 1 3)  & 3501.4  &   2.0 & --0.1 \\
   ( 4 0-- 0)  &  ( 1 0 3)  & 3814.1  & --1.1 & --0.1 \\
\hline					 
 r.m.s.     &       &         &  2.3 & 1.0   \\
\hline
\hline
\end{tabular}
\end{table}

\clearpage
\begin{table}
\centering
\caption[]{Predicted $^{18}$O$_3$ vibrational energy levels 
up to four quanta. The calculations 
correspond to the two parameter sets of Table \ref{par18}. 
Energies are given in cm$^{-1}$.}
\label{vib18}
\vspace{10pt}
\begin{tabular}{|c|c|r|r|}          
\hline
\hline
Local Basis & Normal Basis & $E_{cal}$ & $E_{cal}$ \\  
$(v_1,v_2,v_3)$ & $(\nu_1,\nu_2,\nu_3)$ & & \\
\hline        
  ( 0 0+  2) & ( 0 2 0) &  1324.9 & 1321.2 \\
  ( 0 0+  3) & ( 0 3 0) &  1991.1 & 1981.0 \\  
  ( 0 1-- 2) & ( 0 2 1) &  2276.9 & 2276.9 \\
  ( 1 0+  2) & ( 1 2 0) &  2349.2 & 2348.9 \\
  ( 0 0+  4) & ( 0 4 0) &  2659.6 & 2640.3 \\  
  ( 1 1+  1) & ( 2 1 0) &  2722.6 & 2725.8 \\
  ( 0 3+  0) & ( 1 0 2) &  2919.4 & 2919.3 \\
  ( 0 1-- 3) & ( 0 3 1) &  2926.6 & 2922.4 \\
  ( 1 0+  3) & ( 1 3 0) &  3006.7 & 3001.3 \\
  ( 1 2+  0) & ( 3 0 0) &  3108.6 & 3109.3 \\ 
  ( 1 1+  2) & ( 0 2 2) &  3207.3 & 3210.1 \\
  ( 1 1+  2) & ( 2 2 0) &  3368.6 & 3371.0 \\
  ( 3 0+  1) & ( 1 1 2) &  3539.0 & 3544.0 \\
  ( 2 1-- 1) & ( 2 1 1) &  3636.4 & 3640.8 \\
  ( 2 1+  1) & ( 3 1 0) &  3743.2 & 3747.3 \\
  ( 4 0+  0) & ( 0 0 4) &  3795.9 & 3791.3 \\
  ( 2 2+  0) & ( 2 0 2) &  3921.0 & 3917.5 \\
  ( 3 1-- 0) & ( 3 0 1) &  4017.7 & 4017.0 \\
  ( 2 2+  0) & ( 4 0 0) &  4130.4 & 4130.3 \\
\hline			 
\hline
\end{tabular}
\end{table}

\end{document}